\documentclass[]{jfm}

\usepackage{graphicx}
\usepackage{newtxtext}
\usepackage{newtxmath}
\usepackage{natbib}
\usepackage{hyperref}
\hypersetup{
    colorlinks = true,
    urlcolor   = blue,
    citecolor  = black,
}

\newcommand{\RomanNumeralCaps}[1]
\linenumbers

\usepackage{enumerate}
\usepackage{subfigure}
\usepackage{xcolor}
\usepackage{dsfont} \usepackage{stmaryrd}  \DeclareMathOperator{\hs}{\mathcal{N}}\DeclareMathOperator{\dref}{\mathcal{D}_{ref}} \DeclareMathOperator{\diss}{\mathcal{D}}

\newcommand{\jump}[1]{\left\llbracket #1 \right\rrbracket}
\newcommand{\nsigma}{\mathbf{n}_\Sigma}
\newcommand{\ddt}[1]{\frac{d #1}{dt}}

\newcommand{\transpose}{\mathsf{T}} \newcommand{\ndomega}{\mathbf{n}_{\partial\Omega}}
\newcommand{\ngamma}{\mathbf{n}_\Gamma}
\newcommand{\clspeed}{V_\Gamma}
\newcommand{\Ca}{\text{Ca}} 
\newcommand{\mat}[1]{\mathbf{#1}}

\DeclareMathOperator{\platelength}{L_p} 

\DeclareMathOperator{\sigmawet}{\sigma_w}

\DeclareMathOperator{\visc}{\eta}

\DeclareMathOperator{\Oh}{\text{Oh}}
\DeclareMathOperator{\Bo}{\text{Bo}}

\DeclareMathOperator{\thetaeq}{\theta_e} \DeclareMathOperator{\thetaapp}{\theta_{app}}   

\DeclareMathOperator{\bfriction}{\lambda} \DeclareMathOperator{\betamoffatt}{\beta_{Moffatt}}
\DeclareMathOperator{\betacl}{\beta_{cl}}
\DeclareMathOperator{\betahydro}{\beta_{H}}
\DeclareMathOperator{\betaL}{\beta_{L}}

\newcommand{\hl}[1]{#1}

\title{Bridging the scales in capillary rise dynamics with complexity-reduced models}

\author{M.~Fricke\aff{1}
  \corresp{\email{fricke@mma.tu-darmstadt.de}},
  S.~Raju\aff{1}, E.~A.~Ouro-Koura\aff{1}, O.~Kozymka\aff{1}, J.~De Coninck\aff{2}, Ž.~Tuković\aff{3}, T.~Marić\aff{1}
 \and D.~Bothe\aff{1}}

\affiliation{\aff{1}Mathematical Modeling and Analysis, TU Darmstadt, Peter-Grünberg-Str.~10, 64287 Darmstadt, Germany
\aff{2}Transfers, Interfaces and Processes, Université libre de Bruxelles, Av.~Franklin~Roosevelt~50, Bruxelles, Belgium
\aff{3}Faculty of Mechanical Engineering and Naval Architecture, University of Zagreb, Ivana Lučića 5, 10000 Zagreb, Croatia
}

\begin{document}
\maketitle

\begin{abstract}
Dynamic wetting processes inherently manifest as multiscale phenomena. While the capillary length is typically millimeters, solid-liquid interactions occur at the nanometer scale. These short-range interactions significantly affect macroscopic behaviors like droplet spreading and menisci dynamics. The Navier slip length, determined by liquid viscosity and solid-liquid friction, plays a crucial role in three-phase contact line dynamics. It varies from nanometers (hydrophilic) to microns (hydrophobic). However, resolving it in computational fluid dynamics (CFD) simulations can be computationally expensive. In this study, we propose simplified ordinary differential equation (ODE) models, leveraging local dissipation rates from Stokes flow solutions near the moving contact line, to bridge the nanoscale physics and macroscopic dynamics. Our ODE model accurately predicts the impact of the slip parameter in fully resolved CFD simulations, focusing on capillary rise dynamics.

 \end{abstract}

\section{Introduction}
The \emph{predictive} mathematical modeling of dynamic wetting processes is still a challenge and, under general circumstances, not yet state of the art. This is essentially a consequence of the small-scale physical processes that influence the macroscopic flow significantly. Historically, the story begins with the famous paper by \cite{Huh1971} that points out why the classical no-slip boundary condition for the (two-phase) Navier Stokes equations fails to describe dynamic wetting in the sharp interface modeling framework. In essence, the no-slip boundary condition is kinematically inconsistent with a "material" (i.e.\ passively advected) moving contact line unless a discontinuity of the wall tangential component of fluid viscosity is allowed. This, however, introduces a non-integrable singularity in the dissipation of a viscous fluid. Since then, a variety of different models and physical mechanisms have been proposed that (at least partially) regularize the singularity and make the problem well-posed in an appropriate mathematical setting. For an overview of the large body of literature on the topic, the reader is referred to \cite{Gennes1985,Gennes2004,Shikhmurzaev2008,Bonn2009,Coninck2022}. All the different models introduce additional parameters on top of the common parameters $\rho$ (density), $\eta$ (viscosity), $\sigma$ (liquid-gas surface tension) and $g$ (gravitational acceleration) describing the dynamics of Newtonian liquids under isothermal conditions. One of the most prominent models to resolve the contact line paradox is the Navier slip model, which states that the tangential velocity of fluid particles relative to the solid is proportional to the tangential component of the viscous stress, more precisely
\begin{align}\label{eqn:navier_slip}
-\bfriction \mathbf{v}_\parallel = (\mat{S} \ndomega)_\parallel. 
\end{align}
with the friction coefficient $\bfriction \geq 0$. The viscous stress tensor is denoted as $\mat{S} = 2\eta \mat{D} = \eta(\nabla \mathbf{v} + \nabla \mathbf{v}^\transpose)$, $\ndomega$ is the unit outer normal vector for the solid boundary and $\mathbf{w}_\parallel$ is the parallel component of a vector $\mathbf{w}$ with respect to the solid wall, i.e.\ $\mathbf{w}_\parallel = \mathbf{w} - (\mathbf{w} \cdot \ndomega) \ndomega$. In fact, Equation~\eqref{eqn:navier_slip} can be understood as a balance of the viscous stress force with the sliding friction force $-\bfriction \mathbf{v}_\parallel$. Dividing the equation by the friction coefficient $\bfriction > 0$ yields an equivalent formulation where the ``slip length'' $L:=\eta/\bfriction$ appears as a parameter:
\begin{align}\label{eqn:navier_slip_v2}
\mathbf{v}_\parallel + 2 L (\mat{D} \ndomega)_\parallel = 0.
\end{align}
Here $\mat{D} = \frac{1}{2}(\nabla \mathbf{v} + \nabla \mathbf{v}^\transpose)$ denotes the rate-of-deformation tensor. Equation \eqref{eqn:navier_slip_v2} is usually complemented by the impermeability condition $\mathbf{v} \cdot \ndomega = 0$ describing conservation of mass at the boundary. By now, it is widely accepted that slip exists in real physical systems but the value of $L$ is estimated to be between a couple of nanometers and a couple of microns as reported in \cite{Neto2005}. Because of that, the impact of slip in single-phase flow problems is typically small. For example, the single-phase mass flow rate $Q$ in a channel (or capillary) of radius $R$ at a given pressure gradient can be expressed as (see \cite{Neto2005})
\begin{align}\label{eqn:single_phase_mass_flow_slip}
Q = Q_0 (1 + 4 L/R), 
\end{align}
where $Q_0$ denotes the mass flow rate for no-slip. So in practice, the effect is negligible if $L \ll R$. On the other hand, there is a tremendous difference in the case of a contact line moving through a capillary of the same radius. In fact, the standard theory even breaks down for $L \rightarrow 0$ as shown by \cite{Huh1971}. Hence, it is not an option for the contact line case to pass to the limit $L=0$ as an approximation for $L \ll R$.\\
\begin{figure}
\subfigure[]{\includegraphics[height=0.4\columnwidth]{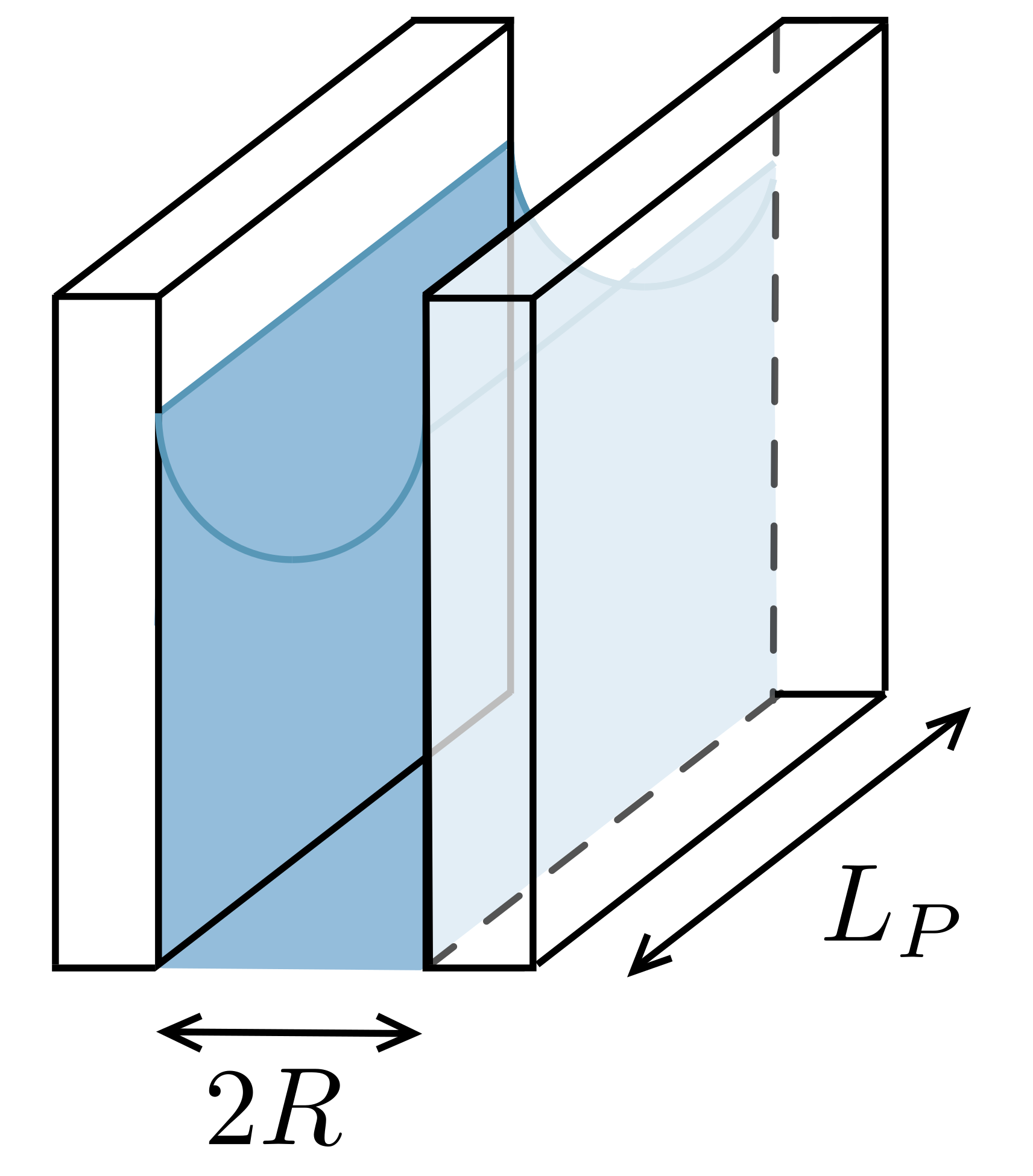}\label{fig:2D_vs_3D}}
\subfigure[]{\includegraphics[height=0.4\columnwidth]{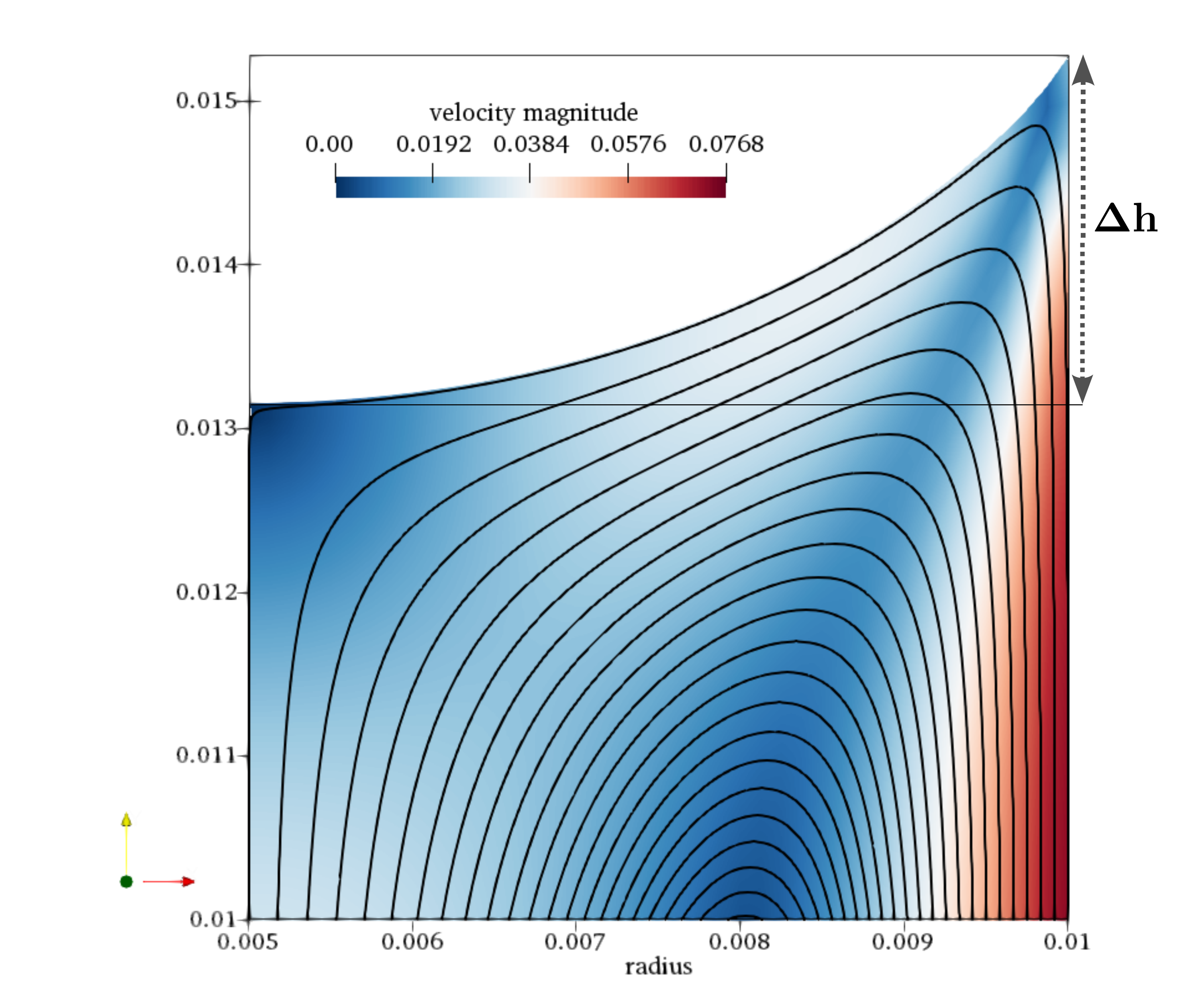}\label{fig:flow}}
\caption{(a): Sketch of geometry with parallel plates. (b): Flow close to the moving contact line (co-moving reference frame). }
\end{figure}
In this work, we focus on the capillary rise problem as one of the prototypical configurations of a dynamic wetting problem. This configuration has been studied intensively over more than hundred years, starting with the seminal work by \cite{Lucas1918}, \cite{Washburn1921} and \cite{Bosanquet1923}. These publications introduced ordinary differential equations for dynamics of the rise height $H$ as a function of time, based on a simplified modeling of the involved forces such as gravity, inertia, capillarity and viscous friction. In Bosanquet's model, the viscous friction is computed only from the Poiseuille flow in the capillary at some distance to the moving interface $\Sigma$. This flow, which actually leads to the relation \eqref{eqn:single_phase_mass_flow_slip}, is only weakly dependent on the slip length. As a consequence, the slip length does not appear as a physical parameter in Bosanquet's model. However, it is evident that the value of the slip parameter plays an important role for the dynamics of the capillary rise. Indeed, it has been shown in numerical simulations by \cite{Gruending2020b}, that lowering the slip length can change the qualitative behaviour of the capillary rise from oscillatory rise (observed for low viscosity liquids) to a monotonic rise regime. For the sake of simplicity, we consider the capillary rise between two parallel plates of length $\platelength$ with a separation $2R$ like sketched in Figure~\ref{fig:2D_vs_3D}. If the length of the plates is much larger than the separation, i.e.\ for $\platelength \gg R$, one can approximate this problem as a quasi two-dimensional capillary rise problem with a capillary of ``radius'' $R$. A generalization of our approach to three dimensions is straightforward. The aim of this study is to describe the capillary rise dynamics in a CFD simulation with a well resolved slip length with a complexity-reduced model. This model is formulated as an ordinary differential equation for the dimensionless rise height $H$ in the form
\begin{align}\label{eqn:complexity-reduced-model}
(H H')' + \frac{\Omega}{1 + \hl{3L/R}} \, H H' + H - 1 = - \beta H'.
\end{align}
This equation can be understood as a generalization of the classical model by Bosanquet, which is contained in \eqref{eqn:complexity-reduced-model} as the special case $\beta=0$ and $L=0$ (see \cite{Fricke2023} for details on the modeling). The parameter $\Omega$ was introduced by \cite{Quere1997} and describes the resistance force arising from the viscous friction in the Poiseuille flow region. Mathematically, it is given as the ratio between the Bond and Ohnesorge\footnote{Note that, in contrast to the standard definition, we define the Bond and Ohnesorge number in terms the specific energy of wetting $-\sigmawet$, which satisfies the Young equation $-\sigmawet=\sigma \cos \thetaeq$. } number, i.e.\
\begin{align}\label{eqn:non-dim-param}
\Omega := \sqrt{\frac{9 \visc^2 \sigma \cos \thetaeq}{g^2 R^5 \rho^3}} = 3 \, \frac{\Oh}{\Bo}, \quad \Bo = \frac{\rho g R^2}{\sigma \cos \thetaeq}, \quad \Oh = \frac{\visc}{\sqrt{R \rho (\sigma \cos \thetaeq)}}.
\end{align}
The key parameter in the present study is the ``generalized friction parameter'' $\beta$, which may be related to different dissipation mechanisms in the dynamic wetting process.

\subsection{Dissipation at the moving contact line}
One important example is the so-called ``contact line friction'' which arises from the microscopic processes within the three-phase contact zone. The Molecular Kinetic theory (MKT) of Blake et al.\ \cite{Blake1969,Blake2002} predicts that, at low velocities of the contact line, the dynamic microscopic contact angle $\theta$ is determined by a balance between a friction force opposing the contact line motion and the out-of-balance Young stress, i.e.\
\begin{equation}\label{eqn:linear-response}
-\zeta \clspeed = \sigma\left(\cos \theta-\cos \thetaeq\right).
\end{equation}
Here, the scalar quantity $\clspeed$ denotes the normal-speed of the contact line measured relative to the solid (with $\clspeed$ > 0 for an advancing and $\clspeed < 0$ for a receding contact line). The parameter $\zeta$, which has the same units as the dynamic viscosity, is the so-called \emph{contact line friction}. Notably, it was stated by \cite{Blake2015} that the contact line friction $\zeta$ and the friction coefficient $\bfriction$ (which determines the slip length via $L=\visc/\bfriction$) should be linked via the width $\delta$ of the three-phase contact zone according to the formula
\begin{align}\label{eqn:friction-link}
\zeta = \bfriction  \delta. 
\end{align}
In fact, this relation suggests that there is only one mechanism of friction between fluid and solid molecules and the contact line friction exists because, physically, the contact line is not sharp but has a finite width determined by the molecular interactions. Inserting \eqref{eqn:linear-response} into the entropy production (see Section~\eqref{sec:modeling} below) leads to a specific form of the parameter $\beta$ given as (with dimensionless contact line friction $\tilde\zeta:=\zeta/\eta$)
\begin{align}\label{eqn:betacl}
\betacl =  \frac{\zeta}{\eta} \Oh = \tilde\zeta \Oh.
\end{align}
Equation~\eqref{eqn:complexity-reduced-model} with $L=0$ and $\beta$ given by \eqref{eqn:betacl} was studied in detail by \cite{Martic2003}. The authors showed that they can describe experimental data by \cite{Quere1997,Quere1999} very well by using $\tilde\zeta$ as a fitting parameter.

\subsection{Hydrodynamic dissipation in the vicinity of the contact line}
It is obvious that the fluid velocity near the contact line (see Fig.~\ref{fig:flow}) has a quite different form than the Poiseuille flow in the channel away from the interface. As pointed out earlier by \cite{Delannoy2019} and \cite{Gruending2020a}, one should account for the viscous friction in the vicinity of the contact line to arrive at a realistic estimate of the overall dissipation. We will refer to this contribution as the ``hydrodynamic dissipation'' in contrast to the contact line dissipation described by the contact line friction and the Poiseuille dissipation described by the parameter $\Omega$. Notably, \cite{Delannoy2019} and Gründing propose quite different models for the hydrodynamic friction parameter. While Delannoy et al.\ argue that the scaling should be logarithmic, i.e.\
\begin{align}\label{eqn:logarithmic-scaling}
\betahydro \sim \ln\left( \frac{R}{L} \right),
\end{align}
the leading-order correction term derived in \cite{Gruending2020a} scales as $R/L$.

\subsection{Goal and structure of this study}
The main goal of this paper is to revisit the theoretical derivation of the hydrodynamic dissipation factor $\betahydro$ and to ``measure'' its dependence on the Navier slip length $L$ by fitting the ODE model \eqref{eqn:complexity-reduced-model} to well-resolved Direct Numerical Simulations. As a framework for our study, we use the two-dimensional capillary rise benchmark problem as described in full detail in \cite{Gruending2020b,Gruending2020b-data}, where all the specifications on the setup and parameters can be found. Since we are focusing on the hydrodynamic dissipation, we suppress the contact line dissipation and apply
\[ \zeta = 0 \quad \Rightarrow \quad \theta = \thetaeq \]
in the numerical simulations. In a first step, we aim at an understanding of the functional dependence of $\betahydro$ on the slip length in the \emph{low} capillary number regime. This means that we will study the capillary rise from an initial height which is already close to the equilibrium state. This allows us to restrict the analysis to the leading-order in the capillary number. An extension of the approach to higher capillary numbers shall be investigated in the future.\\
\\
The remainder of this article is organized as follows: The mathematical modeling for both the continuum mechanical model and the ODE model is briefly revisited in Section~\ref{sec:modeling}. Section~\ref{sec:hydrodynamic-dissipation} shows how the hydrodynamic friction factor $\betahydro$ can be derived in a systematic way. The numerical results are presented in Section~\ref{sec:results}. As a numerical method, we use an Arbitrary Lagrangian–Eulerian (ALE) Interface Tracking method implemented in OpenFOAM, see \cite{Tukovic2012,Gruending2020} for more details. Finally, Section~\ref{sec:conclusions} concludes this work by discussing how the results can open the door to predicting capillary rise dynamics beyond current CFD capabilities. We also briefly discuss how the contact line and hydrodynamic dissipation channels can and should be combined in the future. 
 
\section{Mathematical modeling framework}\label{sec:modeling}
\subsection{Continuum mechanical model}
We consider the sharp-interface and sharp-contact line two-phase Navier Stokes equations for Newtonian fluids in the isothermal case. For small values of the capillary number $\Ca = (\visc \clspeed)/\sigma$, one can expect the ``linear response theory'' \eqref{eqn:linear-response} to be a good approximation to the dynamics of the contact angle. Then, the ``standard model'' based on the Navier slip condition reads as
  \begin{equation}\label{eqn:fundamentals/standard-model-wetting}
 \begin{aligned}
 \partial_t (\rho \mathbf{v}) + \nabla \cdot (\rho \mathbf{v} \otimes \mathbf{v}) - \eta \Delta \mathbf{v} + \nabla p = \rho \mathbf{g}, \quad \nabla \cdot \mathbf{v} = 0, \quad &\text{in} \ \Omega\setminus\Sigma(t),\\
 \jump{\mathbf{v}} = 0, \quad \jump{p \mat{\mathds{1}} - \mat{S}} \nsigma = \sigma \kappa \nsigma, \quad &\text{on} \ \Sigma(t),\\
 \mathbf{v}_\bot = 0, \quad \mathbf{v}_\parallel + 2L (\mat{D} \ndomega)_\parallel = 0, \quad &\text{on} \ \partial\Omega,\\
 V_\Sigma = \mathbf{v} \cdot \nsigma, \quad &\text{on} \ \Sigma(t),\\
 \clspeed = \mathbf{v} \cdot \ngamma, \quad\nsigma \cdot \ndomega = - \cos \theta_0 + (\zeta \clspeed)/\sigma, \quad &\text{on} \ \Gamma(t).
 \end{aligned}
 \end{equation}
We consider the total available energy defined as the sum of the kinetic energy, the surface energies of the free surface $\Sigma(t)$ and the wetted part of the solid $W(t)$, and the gravitational energy, i.e.
\[ \mathcal{E}:= \int_{\Omega(t)} \frac{\rho |\mathbf{v}|^2}{2} \, dV + \int_{\Sigma(t)} \sigma \, dA + \int_{W(t)} \sigmawet \, dA + \mathcal{E}_g. \]

It can be shown mathematically (see \cite{Ren2007,Fricke2019}) that the rate-of-change of the total available enery in the absence of external forces (i.e.\ for $\mathbf{g}=0$) reads as
\begin{align}
\label{eqn:energy_dissipation_continuum}
\ddt{\mathcal{E}} = -2 \int_{\Omega(t)} \eta \mat{D}:\mat{D} \, dV - \int_{\partial\Omega(t)} \frac{\visc |\mathbf{v}_\parallel|^2}{L}  \, dA  - \int_{\Gamma(t)} \zeta \clspeed^2 \, dl.
\end{align}
In the following, we will obtain different ODE models by approximating different contributions to the total dissipation rate \eqref{eqn:energy_dissipation_continuum}. 

\subsection{Complexity-reduced models}
In \cite{Fricke2023}, we derived a generic ODE model for the capillary rise dynamics. It is formulated in the dimensionless variables for the apex height $h$ and time $t$ defined as
\[ H := h/h_0, \quad s:= \frac{t}{\sqrt{h_0/g}}, \]
where $h_0$ is the stationary height. In these variables, the model reads as
\begin{align}\label{eqn:generic_ode}
(H H')' - H'^2 \hs(H') + H - 1 = \frac{1}{H'} \, \frac{\mathcal{D}}{\dref}.
\end{align}
Here, the symbol $\hs$ denotes the function
\begin{align*}
\hs(x) = \begin{cases} 1 & \text{if} \ x \leq 0,\\ 0 &\text{if} \ x > 0. \end{cases} 
\end{align*}
Hence, the second term in \eqref{eqn:generic_ode} is only active while the column is falling, i.e.\ for $H' < 0$. A specific model is obtained from the generic equation \eqref{eqn:generic_ode} by specifying the dissipation mechanisms, i.e.\ the term $\diss$ which describes the total dissipation in the system in dimensional form. It is normalized by the reference dissipation rate
\[ \dref = - 2\platelength \sigmawet \frac{h_0}{\tau} = - 2\platelength \sigmawet \sqrt{h_0 g}. \]
It is straightforward to show that, if the Navier slip condition is applied, the dissipation rate $-2 \int \eta \mat{D}:\mat{D} \, dV$ for the Poiseuille flow region in the two-dimensional setting is given as
\begin{align}\label{eqn:pipeflow-dissipation}
\mathcal{D}_P = - \frac{6  \eta (\platelength/R)  h \dot{h}^2}{1 + \hl{3L/R}} = - \frac{\platelength}{1+3 L/R} \frac{6 \eta h_0^3}{R \, \tau^2} \, H (H')^2.
\end{align}
This yields (see \eqref{eqn:non-dim-param} for the definition of $\Omega$) 
\[ \frac{1}{H'} \, \frac{\mathcal{D}_P}{\dref} = - \frac{\Omega}{1+3 L/R} \, H H'. \]
We keep the dependence on the slip length here because we also study large values of $L$ in the simulations.

\section{Modeling viscous dissipation in the contact line vicinity}\label{sec:hydrodynamic-dissipation}
In order to estimate the viscous dissipation in the vicinity of the contact line, we will use two different asymptotic solutions of the equations in a wedge geometry. At a radial distance $r \gtrsim L$ (``intermediate scale'', Section~\ref{section:moffatt-dissipation}), we use the asymptotic solution for a no-slip boundary condition. On the microscopic scale, we will use an asymptotic solution with Navier slip as a regularization; see Section~\ref{section:dissipation_microscopic_scale}.

\subsection{Viscous dissipation on intermediate scales}\label{section:moffatt-dissipation}
In order to estimate the viscous dissipation for $r \geq L$, we make use of the asymptotic solution of the Stokes flow problem in two dimensions for a no-slip boundary condition given by \cite{Moffatt1964}. In the stream function formulation in polar coordinates ($0 \leq r < \infty$, $0 \leq \varphi \leq \theta$), the solution reads as
\begin{equation}
\psi_1(r, \varphi)=\frac{r  \clspeed[(\varphi-\theta) \sin \varphi-\varphi \sin (\varphi-\theta) \cos \theta]}{\sin \theta \cos \theta-\theta} =: r \clspeed f_1(\varphi).
\end{equation}
From this expression, one can recover the velocity field $\mathbf{\tilde{v}} = \tilde{v}_r \, \mathbf{e}_r + \tilde{v}_\varphi \, \mathbf{e}_\varphi$ (defined in a reference frame co-moving with the contact line) via $\tilde{v}_r = 1/r~\partial_\varphi \psi$ and $\tilde{v}_\varphi = - \partial_r \psi$. We compute the rate of deformation tensor for the Moffatt solution in polar coordinates. It reads as
\begin{align} 
\mat{D}(r,\varphi) = \frac{\clspeed F_1(\varphi)}{2r} \begin{pmatrix}
0 & 1 \\ 1 & 0
\end{pmatrix},
\end{align}
where the function $F_1$ is defined as $F_1(\varphi) = f''_1(\varphi) + f_1(\varphi)$. Now we may perform the integration over the wedge domain $c_1 L \leq r \leq c_2 \Delta h$ and $0 \leq \varphi \leq \thetaapp$ to compute the dissipation rate. Notice that the wedge geometry should be associated with the \emph{apparent contact angle} $\thetaapp$ that is assumed by the interface at a characteristic distance away from the contact line. This characteristic distance is proportional to the difference between apex height and contact line height denoted as $\Delta h$ in the following; see Fig.~\ref{fig:flow}). Since we are interested in the leading-order effect in $\Ca$, we neglect the deviation between apparent contact angle and microscopic contact angle $\theta$. Moreover, we introduce constants $c_1, c_2 > 0$ because the precise domain of validity of the respective solutions is unknown\footnote{To determine $c_1$ and $c_2$ would require a more detailed analyis of the asymptotic matching between the different flow solutions. This is beyond the scope of this study.}. This considerations lead to the following bulk dissipation rate:
\begin{equation}
\begin{aligned}
\mathcal{D}_{\text{Moffatt}} &= - \platelength 2 \eta \int_{c_1 L}^{c_2 \Delta h} \int_0^\theta 2 \left( \frac{\clspeed F_1(\varphi)}{2 r} \right)^2  d\varphi \, r \, dr = -2 \platelength \eta \mathcal{F}_1(\theta) \, \hl{\ln\left( \frac{c_2 \Delta h}{c_1 L} \right) \dot{h}^2}.
\end{aligned}
\end{equation}
Here, $\Delta h > L$ denotes the difference between the contact line height and the apex height (see Fig.~\ref{fig:flow})). The function $\mathcal{F}_1(\theta)$ (which arises from the integration over $\varphi$) is defined as
\begin{align}\label{eqn:geometric_function_moffat}
\mathcal{F}_1(\theta) = \frac{1}{2} \int_0^\theta F_1(\varphi)^2 \, d\varphi = \frac{1}{2} \int_0^\theta (f''_1(\varphi) + f_1(\varphi))^2 d\varphi = \frac{\sin^2 \theta}{\theta - \cos \theta \sin \theta} \geq 0. 
\end{align}
Hence, if $L < \Delta h$, it follows that (we define $\tilde{c}:=c_2/c_1$)
\begin{align}\label{eqn:hydrodynamic_friction_factor}
\boxed{\betamoffatt = - \frac{\mathcal{D}_{\text{Moffatt}}}{(H')^2  \dref} = \mathcal{F}_1(\theta) \, \hl{\Oh \, \ln\left( \frac{\tilde{c} \Delta h}{R} \frac{R}{L} \right)} = \mathcal{F}_1(\theta) \Oh \left( \ln\left( \frac{R}{L} \right) + \ln\left( \frac{\tilde{c} \Delta h}{R} \right) \right) \geq 0.}
\end{align}
If the Bond number is sufficiently small such that the free surface can be approximated by a spherical cap, then the above equation can be further simplified using the relation $\Delta h/R = (1-\sin \theta)/\cos \theta$. The function $\mathcal{F}_1$ may be simplified using the asymptotic relation $\mathcal{F}_1 \sim 3/(2 \theta)$ as $\theta \rightarrow 0$.

\subsection{Viscous dissipation on small scales}\label{section:dissipation_microscopic_scale}
The leading-order asymptotic solution to the Stokes problem with the slip boundary condition at the solid was provided by \cite{Hocking1977} and, recently, in a more detailed variant by \cite{Kulkarni2023}. It is shown that the stream function has the form
\begin{equation}
\psi_L(r, \varphi)=  r \clspeed \left( \frac{r}{L} \mathcal{K}_1 (\varphi) + \left(\frac{r}{L}\right)^2 \mathcal{K}_2 (\varphi) + \left(\frac{r}{L}\right)^3 \mathcal{K}_3 (\varphi) + \dots \right).
\end{equation}
We see that, to leading order for this stream function, the velocity scales like $\mathbf{\tilde{v}} \sim \clspeed \, r/L$. Hence, $\mathbf{\tilde{v}} \rightarrow 0$ in the co-moving reference frame as the contact line is approached and the velocity field is continuous. The velocity gradient scales like $\nabla \mathbf{v} \sim \clspeed/L$ (compared to $\clspeed / r$ for no-slip). Hence, we obtain the following leading-order contribution to the bulk dissipation
\[ \mathcal{D}_L = - \platelength 2 \visc \mathcal{F}_2(\theta) \int_0^{c_1 L} \frac{\clspeed^2}{2L^2} \, r \, dr = - c_1 \platelength 2\visc \mathcal{F}_2(\theta) \clspeed^2.  \]
We observe that the corresponding friction factor is \emph{independent} of the slip length:
\begin{align}
\betaL = -\frac{1}{(H')^2} \, \frac{\mathcal{D}_{L}}{\dref} = \hl{\Oh c_1 \mathcal{F}_2(\theta) \geq 0.}
\end{align}
To leading order, the velocity computed from $\psi_L$ and measured relative to the solid behaves like $|\mathbf{v}_\parallel|^2 = \clspeed^2 (1-r/L)^2$ (for $0\leq r \leq L$) and, hence, we have for the dissipation due to slip (note that $\beta=\visc/L$)
\begin{align}
\mathcal{D}_\parallel = -2 \platelength \int_0^{L} \frac{\visc}{L} \, \clspeed^2 \left(1-\frac{r}{L}\right)^2 \, dr = - 2 \platelength \eta \clspeed^2/3.
\end{align}
The corresponding friction factor reads as
\begin{align}
\beta_\parallel = -\frac{1}{(H')^2} \, \frac{\mathcal{D}_\parallel}{\dref} = \hl{\frac{\Oh}{3}.}
\end{align}
  
\section{Results}\label{sec:results}
\begin{figure}
\subfigure[Rise dynamics CFD for selected values of $R/L$.]{\includegraphics[width=0.5\columnwidth]{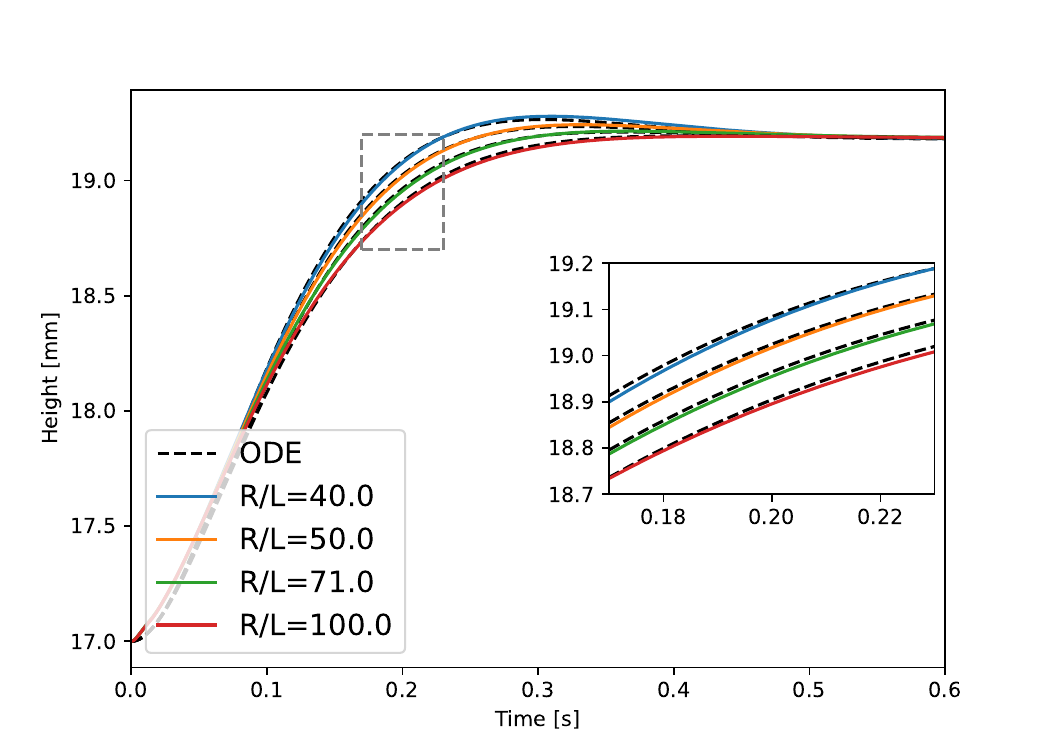}\label{fig:cfd-ode}}
\subfigure[Dependence of $\beta$ on $R/L$.]{\includegraphics[width=0.5\columnwidth]{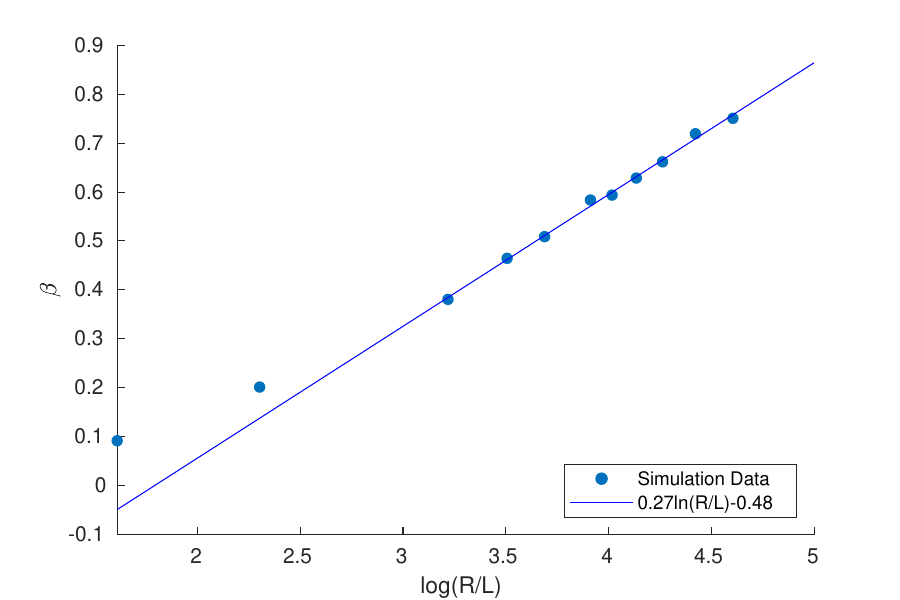}\label{fig:beta-results}}
\caption{Dynamics 2D capillary rise for varying slip length.}\label{fig:results}
\end{figure}
We perform Direct Numerical Simulations using the ALE method of the capillary rise problem with a \emph{fixed} contact angle and a slip length $L$ which is well-resolved by the computational mesh. The setup is identical to the one reported for the ALE method in \cite{Gruending2020b}. However, in contrast to the previous study, we initialize the meniscus close to its equilibrium position to suppress any higher-order effects in the capillary number. We vary the slip length in the interval
\[ \frac{1}{5} \leq \frac{L}{R} \leq \frac{1}{100} \]
to study the dependence on the parameter $\beta$ that is obtained from fitting the CFD simulation data to the generic ODE model \eqref{eqn:complexity-reduced-model}. We build upon the benchmark example for $\Omega=1$ in the paper by \cite{Gruending2020b} because this is an example configuration for which the numerical method is thoroughly validated. The physical parameters are given as
\[ R = 5 \cdot 10^{-3} \, \text{m}, \ \rho = 83.1 \, \text{kg}/\text{m}^3, \ \mu = 0.01 \, \text{Pa} \, \text{s}, \ g = 4.17 \, \text{m}/\text{s}^2 \ \text{and} \ \sigma = 0.04 \, \text{N}/\text{m}.  \]
We suppress the dissipation at the contact line by setting a fixed contact angle $\theta=30^\circ$. Notice that these parameters are designed for the purpose of a numerical benchmark and, therefore, do not correspond to any particular fluid. For the given set of parameters, we have $\Omega = 1$, $\Bo = 0.25$ and $\Oh = 8.34 \cdot 10^{-2}$. Hence, we expect that $\betahydro$ is proportional to $\ln(R/L)$ with \hl{slope $\mathcal{F}_1(\pi/6) \, \Oh \approx 0.23$.} The results from our simulations are shown in Figure~\ref{fig:results}. From Figure~\ref{fig:cfd-ode}, we observe that, as expected, the damping of the initial rise height oscillation is monotonically increasing with $R/L$. A selection of simulation curves are shown together with the ODE solution for a fitted effective friction coefficient $\beta$. The dependence of $\beta$ on $\ln(R/L)$ is reported in Figure~\ref{fig:beta-results}. The numerically obtained proportionality factor is $\approx 0.27$ and, hence, quite close to the estimated value of $0.23$. This indeed shows that the prediction \eqref{eqn:hydrodynamic_friction_factor} is confirmed for the numerical data with a quite good accuracy.

\section{Conclusion and Outlook}\label{sec:conclusions}
This study has shown, as a proof of principle, that the ODE model \eqref{eqn:complexity-reduced-model} is able to systematically capture the hydrodynamic effect of the Navier slip length on the dynamics of capillary rise. This is achieved by employing the modeling framework of complexity-reduced models of capillary rise described in \cite{Fricke2023} which requires an estimate of the dissipation rates in the continuum mechanical approach (see Eq.~\eqref{eqn:energy_dissipation_continuum}). These dissipation rates are estimated based on known asymptotic solutions of the Stokes equation in a wedge geometry. The individual contributions sum up to the total hydrodynamic friction coefficient given as
\begin{align}\label{eqn:summary_hydrodynamic_dissipation}
\betahydro = \mathcal{F}_1(\theta) \Oh \ln\left(\frac{R}{L}\right) + \Oh\left( \frac{1}{3} + \mathcal{F}_1(\theta) \ln\left(\frac{c_2}{c_1}\frac{\Delta h}{R}\right) + c_1 \mathcal{F}_2(\theta)  \right).
\end{align}
Notably, the coefficient of the logarithmic term is completely determined by the product of the Ohnesorge number and a geometric function of the contact angle, which can be computed explicitly (see Eq.~\ref{eqn:geometric_function_moffat}). Predicting the second term in \eqref{eqn:summary_hydrodynamic_dissipation} is more challenging because it involves the computation of the coupling constants $c_1$ and $c_2$. So, in practice, the constant part of $\betahydro$ (with respect to $L$) should be inferred from simulations. From the confirmed logarithmic scaling, it is then possible to predict the $\beta$ parameter for arbitrarily small values of the slip parameter and apply the ODE model beyond the computational limits of the CFD simulation. It has been demonstrated by \cite{Blake2004} that such kinds of models may be useful for the optimization of imbibition and drainage, a problem crucial for applications. However, the model used in the latter work only considers the contact line friction channel of dissipation, which is characterized by
\begin{align}\label{eqn:summary_cl_dissipation}
\betacl = \zeta/\eta \, \Oh.
\end{align}
From a theoretical point of view, it would be very interesting to extend the present approach to a case where the two dissipation mechanisms \eqref{eqn:summary_hydrodynamic_dissipation} and \eqref{eqn:summary_cl_dissipation} are both active at the same time (what is actually expected because of \eqref{eqn:friction-link}). Moreover, it should be noted that the present study only considers the hydrodynamic contribution in the limit $\Ca \rightarrow 0$. For higher values of the capillary number one can expect the apparent contact angle $\thetaapp$, which will become the argument of $\mathcal{F}_1$ in \eqref{eqn:summary_hydrodynamic_dissipation}, to deviate significantly from the microscopic contact angle as explained by \cite{Cox1986}. This effect leads to an interesting non-linear coupling between these two mechanisms and should be studied in detail in the future.

\backsection[Acknowledgements]{}

\backsection[Funding]{We acknowledge the financial support by the German Research Foundation (DFG) within the Collaborative Research Centre 1194 (Project-ID 265191195).}

\backsection[Declaration of interests]{The authors report no conflict of interest.}

\bibliographystyle{jfm}

\end{document}